\documentclass[preprint2]{aastex}
\usepackage{epsfig}
\usepackage{psfig}
\usepackage{graphics}
\usepackage{latexsym}
\newcommand{\hedreiplus}{$^3\mathrm{He}^{2+}$ }
\newcommand{\hedrei}{$^3\mathrm{He}$ }

\newcommand{\hevier}{$^4\mathrm{He}$ }

\newcommand{\fe}{$^{56}\mathrm{Fe}$ }

\newcommand{\ox}{$^{16}\mathrm{O}$ }
\newcommand{\heratio}{$^3\mathrm{He}/^4\mathrm{He}$ }
\newcommand{\feratio}{$\mathrm{Fe}/\mathrm{O}$ }
\newcommand{\mev}{$1\;\mathrm{MeV\;amu}^{-1}$ }
\newcommand{\kev}{$\sim100\;\mathrm{keV\;amu}^{-1}$ }
\begin{document}
\title{ACCELERATION AND ENRICHMENT OF $^{\bf 3}\mathbf{He}$ IN
  IMPULSIVE SOLAR FLARES BY ELECTRON FIREHOSE WAVES}
\shorttitle{$^3\mathrm{He}$ ACCELERATION BY EF WAVES}
\author{G. Paesold}
\affil{Institute of Astronomy\\
       ETH Zentrum, CH-8092 Zurich, Switzerland}
\affil{Paul Scherrer Institute\\
       W\"urenlingen und Villigen, CH-5232 Villigen PSI, Switzerland}
\email{gpaesold@astro.phys.ethz.ch}
\author{R. Kallenbach}
\affil{International Space Science Institute\\
       Hallerstrasse 6, CH-3012, Bern, Switzerland}
\email{reinald.kallenbach@issi.unibe.ch}
\and
\author{A.O. Benz}
\affil{Institute of Astronomy\\
       ETH Zentrum, CH-8092 Zurich, Switzerland}
\email{benz@astro.phys.ethz.ch}
\shortauthors{Paesold et al.}
%
%
\begin{abstract}
A new mechanism for acceleration and enrichment of $^3\mathrm{He}$
during impulsive solar flares is presented. Low-frequency
electromagnetic plasma waves excited by the Electron Firehose
Instability (EFI) can account for the acceleration of ions up to \mev
energies as a single stage process. The EFI arises as
a direct consequence of the free energy stored in a temperature anisotropy
($T^e_\parallel>T^e_\perp$) of the bulk energized electron
population during the acceleration process. In contrast to other
mechanisms which require special plasma properties, the EFI is an
intrinsic feature of the acceleration process of the bulk electrons.
Being present as a side effect in the flaring plasma, these waves can
account for the acceleration of $^3\mathrm{He}$ and $^4\mathrm{He}$
while selectively enhancing $^3\mathrm{He}$ due to the spectral energy
density built up from linear growth. Linearized kinetic theory,
analytic models and test-particle simulations have been applied to
investigate the ability of the waves to accelerate and
fractionate. As waves grow in both directions parallel to the magnetic
field, they can trap resonant ions and efficiently
accelerate them to the highest energies. Plausible models have been
found that can explain the observed energies, spectra and abundances
of $^3\mathrm{He}$ and $^4\mathrm{He}$.
\end{abstract}
\maketitle
%
%
%
%
\section{INTRODUCTION}
Solar flares are commonly divided into two different classes: impulsive
and gradual \citep*{caneetal1986}. The division into these
two categories can be done on the basis of the duration of their soft
X-ray emission \citep*{pallavicinietal1977}. But it is
not only the timescale of the events that justifies the distinction: the
energetic particles observed in space from impulsive flares exhibit
strong abundance enhancements over coronal values
\citep[and references therein]{lin1987,reames1990}.
Impulsive flares are usually
dominated by energetic electrons and are characterized by $^3\mathrm{He}/
^4\mathrm{He}$ ratios at \mev energies that are frequently 3 to 4 orders
of magnitude larger than the corresponding value in the solar corona
and solar wind where $^3\mathrm{He}/^4\mathrm{He}
\sim 5\cdot10^{-4}$. They also exhibit enhanced
$^4\mathrm{He}/\mathrm{H}$ and
$\mathrm{Fe}/\mathrm{C}$ ratios. Although the occurrence of \hedrei and
\fe enrichments are correlated
in impulsive flares, the ratio $^3\mathrm{He}/\mathrm{Fe}$ shows huge
variations as observed by \citet*{masonetal2000}. This
suggests, that different mechanisms are responsible for the
acceleration of the two species. Gradual flares usually have large
energetic proton fluxes, small $^4\mathrm{He}/
\mathrm{H}$ and do not show large $^3\mathrm{He}/
^4\mathrm{He}$ or $\mathrm{Fe}/\mathrm{C}$ enhancements in the
energetic particles, although approximately $5\%$ admixture of
suprathermal remnant particles from impulsive flares have been
observed in gradual events by \citet{tylkaetal2001}. The standard
interpretation for these
observations is that the energetic particles in impulsive events
origin in the energy release region on the sun while the energetic
particles in gradual events are accelerated via shocks, either coronal
or interplanetary \citep{lin1987,luhnetal1987}.

Abundance ratios therefore are a valuable diagnostics for the
flaring plasma itself and in particular the specific acceleration
mechanism for the energetic particles. The selectivity of the
mechanism, especially for \hedrei and
\hevier indicates resonant processes
such as gyroresonant interaction of plasma waves with the
ions. Theoretical ideas therefore focus on the unique
charge-to-mass ratio of \hedrei which allows
it to be selectively pre-heated or accelerated via gyroresonance.

A well-known theory of the initial set among theories for
\hedrei enhancement was published by \citet{fisk1978},
explaining the preferential acceleration of the ions by
electrostatic ion cyclotron (EIC) waves at a frequency in the
vicinity of the gyrofrequency of \hedrei.
The waves are excited by an electron current and interact with
\hedrei via cyclotron resonance. A large enhancement of
$^4\mathrm{He}/\mathrm{H}$ is required in the ambient plasma
for this instability to excite waves above the \hevier gyrofrequency.

More recently a theory was suggested by \citet{temerinroth1992}
who accounted for the preferential
\hedrei acceleration proton electromagnetic ion cyclotron
($\mathrm{H}^+$ EMIC) waves.
These waves are driven unstable by non-relativistic (keV range) electron
beams and their frequencies lie at around the \hedrei
gyrofrequency at almost perpendicular propagation. In
\citet{temerinroth1992},
auroral observations of keV electron beams and $\mathrm{H}^+$~EMIC waves
are taken as experimental evidence that $\mathrm{H}^+$ EMIC waves also may
acquire a substantial fraction (order of few percent) of the electron
beam energy under coronal conditions. At the Sun, plasma emission from
tens of keV electron beams on open magnetic field lines is the
explanation for type III radio emission with its equivalent, the
U-bursts, on closed field lines (in loops) at beam energies of
order keV. In order to excite $\mathrm{H}^+$~EMIC waves fulfilling the
requirements of this model, electron beams of much higher density than
the observed ones have to be postulated \citep{millervinas1993} with
no direct observational evidence.
Moreover, it is difficult to explain from a theoretical point of
view that the free energy in the electron beam is transferred to
the $\mathrm{H}^+$~EMIC waves and not to the much faster growing
($\sim 4$ orders of magnitude in growth rate) electron plasma waves.

In this work an alternative model for acceleration of
\hedrei and its enhancement over \hevier is presented.
The approach is different from the models described above. While
the models mentioned above postulate rather special plasma properties
($^4\mathrm{He}/\mathrm{H}$ overabundance in the pre-flaring plasma, dense
low-energy beams) in order to produce the required plasma waves, the
model presented herein explains the
unique overabundance of \hedrei by plasma waves
excited as an intrinsic feature of the electron acceleration
process itself.
Parallel propagating, lefthand polarized electromagnetic
waves driven by the Electron Firehose instability (EFI) can account for
\hedrei acceleration via gyroresonant interaction. As suggested in
\citet{paesoldbenz1999} such Electron Firehose (EF) waves are excited
by anisotropic electron distribution functions
($T_\parallel>T_\perp$) that occur in the course of
the acceleration process of bulk electrons in solar flares.

Although the EF waves are not narrowbanded around the gyrofrequency of
\hedrei as the $\mathrm{H}^+$~EMIC proposed
by~\citet{temerinroth1992}, selectivity of the process is achieved by
the natural profile of the spectral wave energy. No additional assumptions
besides the electron anisotropy are needed,
and the model can be embedded as an intrinsic feature in acceleration
scenarios such as transit-time damping, the currently most popular
stochastic acceleration model.

In \S~\ref{idea} the basics of the new acceleration scenario are
described. The
properties of the EF waves under coronal conditions
are presented in \S~\ref{props}. Analytical and numerical
results on the heating rates are presented in \S~\ref{heating}
and \S~\ref{sim}. The mechanism for enhancement of \hedrei over \hevier is
described in \S~\ref{enhancement}, and heavier ions are discussed
in \S~\ref{heavy}. \S~\ref{conc} concludes this work.
%
%
\section{BASIC IDEA}
\label{idea}
Among the most promising scenarios for accelerating electrons to
observed energies in impulsive solar flares is transit-time damping
acceleration~\citep{fisk1976,stix1992}, the magnetic analogon of
Landau damping. The following scenario was presented by
\citet{millerlarosamoore1996}: Electrons are accelerated
from thermal to relativistic
energies by resonance with low-amplitude fast-mode waves having a
continuous broadband spectrum. The magnetic moment of the particle
interacts with the parallel gradient of the magnetic field as in
the well known Fermi
acceleration~\citep{fermi1949,davis1956}. Contrary
to the classic Fermi process, transit-time damping involves small-amplitude
magnetic compressions. While Fermi acceleration is the result of
large numbers of particles being reflected by randomly moving magnetic
compressions, transit-time damping is a process of rather resonant nature.
In the limit of very small amplitudes, only particles of
the same parallel speed as the wave phase velocity can be reflected
by the magnetic compressions, i.e. $v_\parallel=\omega/k$ which is the
Landau resonance condition. Interaction with a wave changes the
particle's parallel speed and therefore allows it to interact with
another wave of the continuous spectrum. In the average this process
results in stochastic acceleration.

The wave spectrum is assumed to origin from cascading fast-mode
waves, initially excited at very long wavelengths as a direct output
of the primary energy release in impulsive flares. The cascade
channels the released energy through an inertial region to $k$-values
small enough to accelerate electrons out of the thermal background
population.

Other acceleration scenarios operating in impulsive solar flares have
been proposed and can roughly be divided into three categories: Shock
acceleration, acceleration by parallel electric fields and stochastic
acceleration by MHD turbulence including the model described above. A
detailed review can be found in \citet{milleretal1997} and references
therein.

Even though the nature of the actual acceleration mechanisms is
unclear, all the above mechanisms that can account for accelerating the
bulk of electrons up to the observed energies of $\sim
20\;\mathrm{keV}$ have an important feature in
common: Particles are preferentially
accelerated in parallel direction with respect to the background
magnetic field.
Parallel dc electric fields trivially accelerate only along the
magnetic field while stochastic scenarios as transit-time damping
act via small amplitude magnetic mirroring which is only capable of
transferring energy in parallel direction if no additional scattering
mechanism is provided ~\citep{lentersmiller1998}. Works of
e.g. \citet{wu1984} and 
\citet{leroymangeney1984} describe the parallely directed energization of
electrons at the earths bow shock via shock drift acceleration with
quasi perpendicular shocks.

The distribution function of the accelerated particle
species in velocity space therefore is expected to become more and
more anisotropic in the course of the energization process.
While the perpendicular temperature remains virtually constant during
the acceleration, the parallel velocity of the particles increases.
If the energization in parallel direction is from a
thermal level of some 0.1 keV up to 20 keV or more, the anisotropy in
parallel direction is substantial. The plasma therefore is modeled
by a bi-maxwellian with different temperatures in parallel and
perpendicular direction.
In doing so it is assumed that no efficient scattering mechanism is
present that could maintain isotropy on acceleration timescales.
Although it is not expected that acceleration retains a bi-maxwellian
distribution function, it is a good
approximation in order to describe propagation properties of plasma
waves. As shown in \citet{paesoldbenz1999} the EFI can occur in such a
situation and give rise to EF waves. The parallel EF waves are purely
transverse electromagnetic and lefthand circularly polarized. The
sense of polarization and can change in certain $k$ ranges if a similar
anisotropy of the protons is assumed.
In the
following the EF waves are lefthand polarized if not mentioned
otherwise. A representative dispersion plot is depicted in
Figure~\ref{fig:1}.

Taking into account the uncertainties in the acceleration region,
including possible pre-heating mechanisms, reasonable pre-flaring
plasma conditions of an impulsive flare range within
$B_0\approx100-500\;\mathrm{G}$ for the background magnetic field,
$n_\alpha\approx10^9-10^{11}\;\mathrm{cm}^{-3}$ in number density
and about $T_\alpha\approx 10^6-10^7\;\mathrm{K}$ for the proton
and electron temperature \citep{pallavicinietal1977}.
For the numerical example herein the following pre-flaring
plasma parameters are chosen: $n_e=n_p=5\cdot 10^{10}\;\mathrm{cm}^{-3},
T^e_{\perp,\parallel}=T^p_{\perp,\parallel}=1\cdot10^7\;\mathrm{K},
B_0=100\;\mathrm{G}$.

Under these exemplary conditions the EF waves propagate below about $3\cdot
\Omega_\mathrm{H}$.
Several ion gyrofrequencies are indicated in Figure~\ref{fig:1} and
they lie well within the wave spectrum.
Resonant acceleration occurs if
the condition $\omega-k_\parallel v_\parallel-l\Omega/\gamma=0$ is
satisfied. Here, $v_\parallel$ and $\gamma$ are the parallel
particle speed and Lorentz factor, and $\Omega$ is the gyrofrequency
of the according particle. This condition is well satisfied
for ions like \hedreiplus and $^4\mathrm{He}^{2+}$. The model
presented herein assumes acceleration of the ions via the most
effective gyroresonance at $l=1$, the so called {\em cyclotron}
resonance.

Due to the symmetry of the distribution function with respect to
$v_\parallel$, the dispersion of the EF waves is the same for the
transition from $\mathrm{k}\rightarrow-\mathrm{k}$ while keeping
lefthand polarization and $\omega>0$ \citep{hollwegvolk1970}.
The resulting wavefield
from the EFI therefore consists of waves propagating parallel to the
magnetic field (positive k-branch) as well as anti-parallel
(negative k-branch). This property of the EF waves is of
special interest and crucial for efficient acceleration:
A unidirectional wavefield of electromagnetic waves always exhibits a
force parallel to the background magnetic field pushing the
particle out of resonance, limiting the acceleration time and, hence,
the reachable energies. While other models
(e.g. \citealt{rothtemerin1997}) solve the problem by imposing a
background magnetic field geometry, i.e. a field gradient, to force
the particle to regain resonance, this is not
necessary for the case of the EF waves: resonant acceleration by
counterpropagating electromagnetic waves leads to oscillating parallel
forces and does not drive the particle permanently out of
resonance. In contrary they naturally 'trap'
the particle in the wavefields by bouncing it from resonance
with the positive branch to resonance with the negative branch of
the wavefield and vice versa. This ping-pong trapping strongly
enhances the total resonant interaction time of the particle with the
wave and allows significant acceleration in perpendicular
velocity. The mechanism works even for small wave fields
where 'normal' wave-particle trapping is not efficient.
Similar situations have been investigated
by e.g. inferring counterpropagating Alfv\'en waves to the problem of
solar flare proton acceleration~\citep{barbosa1979} and cosmic-ray
acceleration ~\citep{skilling1975}.
This acceleration can be regarded as a special case of wave turbulent
stochastic acceleration with waves explicitly counterpropagating in
one dimension. 

The differences in enrichment of \hedrei and \hevier is the result of varying
growth of the waves at different frequencies. As the waves grow from a
thermal level to a saturated state, the spectral energy density
develops according to the growth rates from linear theory depicted in
Figure~\ref{fig:1}.
Since the waves are excited nonresonantly by the electrons, the
saturated wave energy spectrum can be approximated from the
profile of the linear growth rate. This is not true for resonantly
excited instabilities, where wave modes can still grow while others have
already saturated. If each mode is in resonance with only the
particles fulfilling the resonance condition, some modes can grow
longer than others and the saturated energy spectrum of the waves has
to be determined by other methods.
However, if the instability is nonresonant as the EFI, each mode
grows according to the total free energy available (the driver
is basically a pressure anisotropy of the electrons) and the whole
electron population
contributes to the growth of each mode. All modes therefore
saturate at the same time, namely when the pressure anisotropy drops below
the threshold, and the spectral energy is frozen at that
time. The wave dispersion is altered by the
decreasing anisotropy as result of the erosion of the particle
distribution. However, a former analysis shows~\citep{paesoldbenz1999}
that this does not severely alter the slope in the
growth rate around \hedrei an $^4\mathrm{He}$.

Although the differences in growth rates are quite
small in the region of interest (see Fig.~\ref{fig:2}), the
differences in wave energy will become significant after several
growth times $\tau=1/\gamma$. As a rough estimate one obtains, by
assuming constant growth rates in time, that two modes with initial
growth rates $\gamma_\mathrm{A}=\gamma_\mathrm{Max}$ and
$\gamma_\mathrm{B}=0.9\;\gamma_\mathrm{Max}$ and equal energy content
have a ratio in energy of $W_B/W_A\sim0.1$ after a time $\tau=\gamma_A
t$ of only $\approx 10$. This estimate illustrates the rather
strong influence of growth rate on the resulting wave energy level
and, hence, on the selectivity of cyclotron acceleration due to the
$\gamma$ profile in wave frequency.
The analysis in~\citet{paesoldbenz1999} shows that the positive slope in
the growth rate profile around the \hedrei and \hevier gyrofrequency
that is needed to selectively enhance \hedrei
is a stable feature of the EFI and not very sensitive to changes in the
plasma parameters for solar pre-flaring conditions.

In the following it will be established that the mechanism meets the
following requirements for the observed \hedrei enrichment:

{\em (i)} The accelerator has to energize the ions above $\sim
1\;\mathrm{MeV\;amu}^{-1}$ on timescales of $1\;s$.

{\em (ii)} \heratio$>0.1$ above energies of about \mev.

{\em (iii)} A time integrated total of about $10^{31}$ $^3\mathrm{He}$
nuclei is required to account for the observed particle
fluxes~\citep{reamesetal1994}.

{\em (iv)} The \hedrei spectrum is harder than the \hevier spectrum in
the range of  $0.4-4.0\;\mathrm{MeV\;amu}^{-1}$ \citep{moebiusetal1982}.

{\em (v)} \hedrei exhibits a turnover in the particle energy spectrum
at around a few \kev~\citep{masonetal2000}.
%
%
\section{ACCELERATION}
\label{acc}
%
%
\subsection{Electron Firehose Wave Properties}
\label{props}
\begin{figure}[ht]
 \plotone{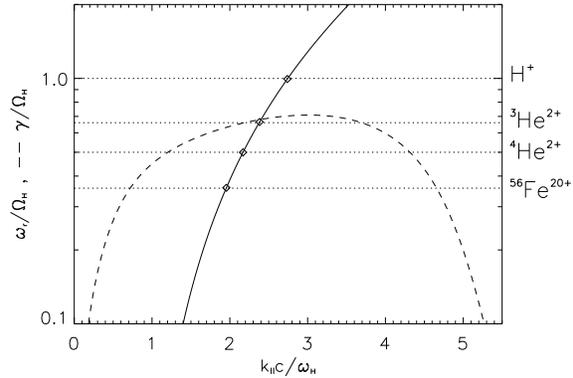}
 \caption[]{Dispersion relation of Electron Firehose waves calculated
 from linear theory. The frequency
   $\omega_r$ (solid curve) and growth rate $\gamma$ (dashed curve) of
   the EF waves are shown normalized to the proton gyrofrequency as function of
   parallel wavenumber $k_\parallel$ normalized to the inertial length
   of the protons.
   The plasma parameters are $T_\parallel^e/T_\perp^e=15$,
   $T_\parallel^p/T_\perp^p=2$, $T_\perp^e=T_\perp^p=1\cdot10^7\;K$,
   $n_e=n_p=5\cdot10^{10}\;\mathrm{cm}^{-3}$
   and $B_0=100\;\mathrm{G}$. The sign convention is
   that $k>0$ and $\omega>0$ mean lefthand
   circularly polarized;
   $\gamma>0$ refers to growing modes. The dotted horizontal lines
   indicate the gyrofrequency of the according ion species.}
 \label{fig:1}
\end{figure}
\citet{parker1958} pointed out that a magnetized plasma with
a pressure anisotropy in parallel direction to the magnetic field can
become unstable to low frequency Alfv\'en waves. This instability is
known as the Firehose instability and is of
completely non-resonant nature: neither the electrons nor the protons
are in resonance with the excited waves. An extension of this
instability to higher frequencies was presented by \citet{hollwegvolk1970}
and \citet{pilippvolk1971}.
This branch of the instability has been termed the Electron Firehose
Instability. Here, the bulk of the protons is resonant with the
waves which are non-resonantly excited by the electrons.
The electrons are anisotropic and drive the waves while the protons
carry the wave. A typical dispersion relation of the EF waves is
displayed in Figure~\ref{fig:1}. The dispersion of the EF waves is
computed by IDLWhamp \citep{paesold2002}, an easy to use IDL interface
to the WHAMP code originally developed by \citet{roennmark1982}. The
code provides the user with the full solution of the dispersion equation
in linearized kinetic theory.

The influence of the presence of other majority ions such as
\hevier on the EF wave dispersion is negligible.
For a plasma consisting of 5\% \hevier and
95\% $\mathrm{H}$ the wavenumber of maximum growth and the frequency at
maximum growth is shifted by values of order of 1\% with
respect to the values of a pure $\mathrm{H}$ plasma.
Ions other than $\mathrm{H}^{+}$ have therefore been omitted in the
following when computing the dispersion relation of the EF waves.
\begin{figure}[ht]
 \plotone{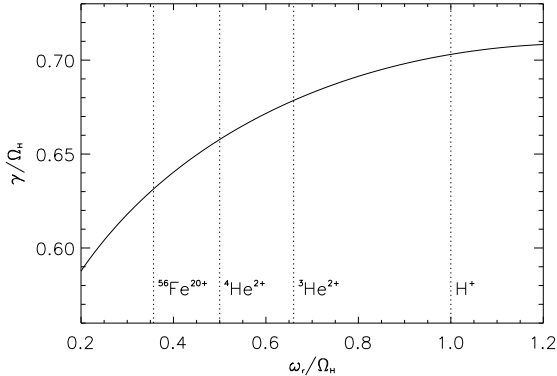}
 \caption[]{Growth rate $\gamma / \Omega_H$ vs. frequency
   $\omega_r / \Omega_H$ for the same plasma as in Figure~\ref{fig:1}.
   Vertical dotted lines indicate the gyrofrequencies of the according
   ions.}
 \label{fig:2}
\end{figure}
\begin{figure}[ht]
 \plotone{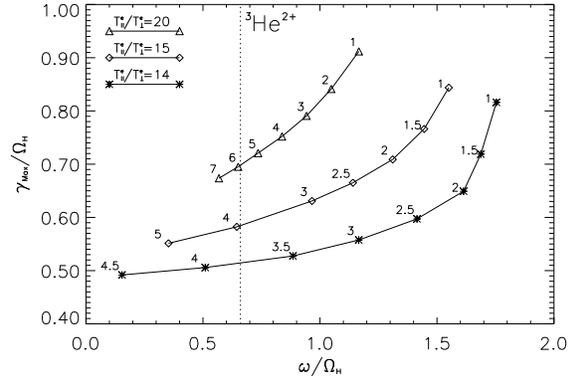}
 \caption[]{Maximum growth rate $\gamma_\mathrm{Max} / \Omega_H$
 vs. frequency $\omega / \Omega_H$. For each of the three lines the
 electron anisotropy is constant and the proton anisotropy is
 varied. The most upper line (triangles) depicts
 $T_\parallel^e/T_\perp^e=20$, the middle line (diamonds)
 $T_\parallel^e/T_\perp^e=15$ and the bottom line (asterisks)
 $T_\parallel^e/T_\perp^e=14$. The small numbers refer to the
 according values of $T_\parallel^p/T_\perp^p$. The vertical dotted
 line indicates the \hedreiplus gyrofrequency.}
 \label{fig:3}
\end{figure}
Due to the assumed anisotropy of the
protons the mode is righthand polarized ($\omega_r<0$) at small
values of k in Figure~\ref{fig:1}. According to \citet{hollwegvolk1970}
the condition for the change of polarization sense to occur is an
additional proton anisotropy $T_\parallel^p/T_\perp^p\geqslant 2$.
Such a proton anisotropy could result e.g. from transit-time
acceleration as described for the electrons in \S~\ref{idea}.
Note, that the proton anisotropy
is not a necessary condition for instability. It has been taken into
account herein only for the sake of generality.
The dependence of the maximum
growth rates on proton and electron anisotropy can be seen in
Figure~\ref{fig:3}. The proton anisotropy has been varied from 1 to larger
values for three values of the electron anisotropy. Only lefthand
modes have been taken into account. With increasing proton anisotropy
the frequency at maximum growth approaches and finally crosses the
\hedrei gyrofrequency. With decreasing electron anisotropy less
anisotropic protons are needed to shift the most growing mode to the
\hedrei gyrofrequency.
This shows that for reasonable anisotropies the slope in the growth
rate profile in the vicinity of the gyrofrequency of \hedrei is always
positive, such that waves around 
the \hedrei resonance grow faster than waves around the \hevier
gyrofrequency. This is a very stable feature of the EF wave spectrum
and can be expected, 
once above the instability threshold, for a large range of plasma parameters
that allow instability and reasonable anisotropies of protons
and electrons \citep[for a detailed analysis of EFI thresholds and spectra in
dependence on plasma parameters see][]{paesoldbenz1999}.

An additional mode can be found at oblique
propagation angles \citep{paesoldbenz1999,lihabbal2000}.
For the reported examples, the oblique mode grows faster than the
parallel mode. In addition to purely growing
waves, the oblique mode contains waves that may resonate with coronal
ions. Relevant for the flare application and mode comparison is the
instability threshold and the non-linear evolution. Also unknown in
this context are the effects of ion anisotropy, power-law electron
distribution and other parameters. Thus 
we concentrate here on the well known parallel mode and leave the
exploration of the oblique mode to future work.
%
%
\subsection{Heating Rates}
\label{heating}
%
%
\subsubsection{Unidirectional Propagation}
\label{uni} The ion's energy change is dominated by the resonant interaction with the wave's
electric field, which is decomposed into its parallel component $E_\parallel =
\hat{E}_\parallel \cos\Psi$, where $\Psi=k_\parallel z+k_\perp x-\omega t$ is the wave's phase
at the location of the ion, and two circularly polarized components
$E_\pm=\hat{E}_\pm(\cos{\Psi},\mp\sin{\Psi})$ with amplitudes $\hat{E}_\pm=(E_x\pm E_y)/2$. In
the Fourier domain we decompose the wave into a set of monochromatic waves separated by a
frequency interval $\Delta \omega$. In one of these monochromatic waves with amplitudes
$\hat{e}_\parallel$ and $\hat{e}_\pm$ and frequency $\omega$, the ion gains or loses energy
according to
\begin{eqnarray}
\dot{W}&=&q v_\perp \hat{e}_\pm\cos{\phi_\pm}+ qv_\parallel
\hat{e}_{\parallel} \cos{\Psi} ~ ,\label{energy}
\end{eqnarray}
where the relative angle $\phi_{\pm}=\theta \pm \Psi$ between the
wave electric field and the perpendicular velocity vector
$(v_x,v_y)=v_\perp(\cos{\theta},\sin{\theta})$ of the ion, moving
with the instantaneous gyrophase $\theta(t)$.

The ion may be driven into or out of cyclotron resonance due to
changes in the ion's parallel speed, which in turn depends on the
wave's electric and magnetic field. The latter is obtained via
Faraday's law $\dot{\mbox{\boldmath $B$}}= - \mbox{\boldmath
  $\nabla$}\times\mbox{\boldmath $E$}$ 
as $(1/\omega)$ $[\pm(k_\parallel\hat{e}_\pm)\sin{\Psi},$
$(k_\parallel\hat{e}_\pm-k_\perp\hat{e}_\parallel)\cos{\Psi},$
$\mp(k_\perp\hat{e}_\pm)\sin{\Psi}]$, so that the parallel acceleration is
\begin{eqnarray}
\label{parallel}
m\dot{v}_\parallel&=&q\left[\left(\frac{k_\parallel
v_\perp}{\omega}\right)\hat{e}_\pm\cos{\phi_\pm}+\right.\nonumber\\
&+&\left.\left(1-\frac{k_\perp v_\perp\cos{\theta}}{\omega}
\right)\hat{e}_\parallel\cos{\Psi}\right] ~.
\end{eqnarray}
A possible mirror force due to a background magnetic field
gradient has been omitted since $B_0$ is assumed to be homogeneous
herein.

The instantaneous ion angular frequency, $\dot{\theta}$, is given by
the ratio of the force on 
the ion, perpendicular to the velocity and the background magnetic
field, and the moment 
perpendicular to the magnetic field. Considering only one
monochromatic wave at frequency $\omega$, the phase equation
$\dot{\phi}_\pm= \dot{\theta}\pm\dot{\Psi}$ is given by 
\begin{eqnarray}
\dot{\phi}_\pm &\approx& -\Omega_i \pm(k_\parallel v_{\parallel}-
\omega+k_\perp v_{\perp}\cos{\theta})\nonumber\\
&-&\frac{q\hat{e}_\pm}{mv_{\perp}}\left(1-\frac{k_\parallel v_{\parallel}}{\omega}\right)\sin{\phi_\pm} ~,
\label{full_phase}
\end{eqnarray}
where $\Omega_i$ is the gyrofrequency of the ion $i$.

In case of the EF waves, where $k_\perp=0$, $\hat{e}_\parallel=0$,
and $\hat{e}_{+}=0$, Eqs.~\ref{energy} and~\ref{parallel} simplify
to
\begin{eqnarray}
\dot{W}&=&q\hat{e}_{-} v_{\perp} \cos{\phi_{-}} ~,\label{acc_en}\\
\dot{v}_{\perp} &=& \frac{q\hat{e}_{-}}{m} \cos{\phi_{-}} \left(1 -
\frac{k_\parallel
v_{\parallel}}{\omega} \right) ~, \label{acc_perp} \\
\dot{v}_{\parallel}&=&\frac{q\hat{e}_{-}}{m}\frac{k_\parallel
v_{\perp}}{\omega}\cos{\phi_{-}}\label{acc_par} ~.
\end{eqnarray}

For a particle $i$ in resonance with a monochromatic EF wave the
frequency mismatch parameter
$\xi=\Omega_i+k_{\parallel} v_{\parallel}-\omega$ vanishes. The
righthand side of
Equation~\ref{full_phase} therefore reduces to a nonlinear
pendulum equation for the phase $\phi_{-}$. In resonance $\phi_{-}(t)$
can have a stable solution
when it stays close to 0. In this case
the particle is accelerated in parallel direction until it is expelled
from resonance (Eq.~\ref{acc_par}). For particles not in resonance,
$\phi_{-}$ changes rapidly and no energy is gained in the time average.

In a set of $N$ monochromatic waves, numbered by $j =1,\dots N$ with equal amplitudes
$\hat{e}_{-,j}= E_\mathrm{rms} / \sqrt{N}$ and equally spaced by a frequency $\Delta \omega$
with $\delta \omega = N \Delta \omega$, the ion's perpendicular motion has a phase $\phi =
\theta - \Psi_j$
with respect to each mode. This gives a set of $N$ phase equations
\begin{eqnarray}
\dot{\phi}_j&=&-\Omega_i - k_{\parallel,j} v_{\parallel} - \omega_j
\nonumber \\ & - & \sum_{l = 1}^{N} \frac{q E_\mathrm{rms}}{m
v_{\perp} \sqrt{N}} \left(1-\frac{k_{\parallel,l}
v_{\parallel}}{\omega_l}\right) \sin {\phi_l } , \label{set_phase}
\end{eqnarray}
and Equations~\ref{acc_perp},~\ref{acc_par} have to be summed over the contributions of the
$N$ modes.

The equations of motion are in general not integrable because of the
$\sin{\phi_l}$ terms in Equation~\ref{set_phase}.
However, a maximum heating rate and the
typical time after which an ion is driven out of resonance by the
mean parallel force of the uni-directional propagating waves are
estimated. The latter limits the total energy gain of the ion.

Within a time interval $\tau$, the energy gain of the ion is mainly
due to the wave modes in a bandwidth $2\pi/\tau$ around the frequency
$\omega\approx\Omega_i+k_\parallel v_\parallel$. The wave may be
described by a mode spacing $\Delta \omega=2\pi/\tau$, yielding
$\hat{e}_{-}\approx E_{rms}/\sqrt{\delta\omega\tau/2\pi}$. At best,
the ion stays in resonance with one mode $j_0$ such that
$\phi_{j_0}\approx0$ over the whole time period $\tau$. This gives,
using Equation~\ref{acc_perp},
\begin{eqnarray}
v_\perp(t+\tau)-v_\perp(t)\approx\sqrt{2\pi}\frac{q E_{rms}\tau}{m \sqrt{\delta\omega\tau}}~,
\end{eqnarray}
and if $\tau$ is sufficiently large so that $v_\perp(t)\ll
v_\perp(t+\tau)$, the heating rate per mass is at most
\begin{eqnarray}
\label{heat_perp}
v_\perp\dot{v}_\perp &\approx& \frac{2\pi q^2 E_{rms}^2}{m^2 \delta\omega}\\
\frac{\dot{W}_\perp}{m}&\approx& 6.1\left[\frac{\mathrm{Mev}}{\mathrm{amu\;s}}\right]\times\\
&\times&\left(\frac{E_{rms}}{100\;\mathrm{V\;m}^{-1}}\right)^2\!\!
\left(\frac{10^6\mathrm{rad\;s}^{-1}}{\delta\omega}\right)\left(\frac{\mathrm{Q}}{\mathrm{A}}\right)^2,\nonumber
\end{eqnarray}
independent of the length of the time interval $\tau$. Thus,
Equation~\ref{heat_perp} predicts a linear increase of energy in
time. The ratio Q/A is the charge-to-mass ratio of the ion in atomic
units.

The resonant heating is limited to the time, during which the ion is
in resonance with one of the wave modes. The ion dynamics are
dominated by the timescales for parallel and perpendicular
acceleration
\begin{eqnarray}
\tau_\parallel &= & \frac{v_\parallel}{\dot{v}_\parallel}=\frac{m
v_\parallel\omega}{q\hat{e}_{-} k_\parallel v_\perp}\\
\tau_\perp &= & \frac{v_\perp}{\dot{v}_\perp}
=\frac{m v_\perp}{q\hat{e}_{-}}\frac{1}{1-k_\parallel v_\parallel/\omega}~.
\end{eqnarray}

The initial conditions are usually $v_{\perp 0}\approx v_{\parallel
  0}$ with $k_\parallel v_\parallel\ll \delta \omega$ and $\tau_{\perp
  0}/\tau_{\parallel 0}<1$. With increasing $v_\perp$, the timescale
  $\tau_\perp$ increases and the timescale $\tau_\parallel$
  decreases. At time $t\approx t_r$, the ion eventually reaches a
  parallel speed $|v_\parallel|$ of order $\delta\omega/k_\parallel$
  and is driven out of resonance. The acceleration timescale at $t_r$ is
\begin{eqnarray}
\tau_{\parallel,r}&\approx&\frac{m\delta\omega
\omega\sqrt{\delta\omega\tau_{\parallel,r}}}{\sqrt{2\pi}q
  E_{rms}k_\parallel^2 v_\perp(t_r)}\nonumber\\
&\approx& \frac{m^2\omega^2\delta\omega^3}{2\pi q^2 E_{rms}^2
  k_\parallel^4}\frac{1}{v_\perp^2(t_r)}\label{t_r}\\
&\approx&  \frac{m^4\omega^2\delta\omega^4}{(2\pi)^2 q^4 E_{rms}^4
  k_\parallel^4}\frac{1}{t_r}~.\nonumber
\end{eqnarray}
Estimating $t_r$ shows that the total time of
the resonant acceleration process is not much longer than
$\tau_{\parallel,r}$, and thus from Equation~\ref{t_r} follows
\begin{eqnarray}
t_r\!\!&\approx&\!\!\frac{m^2 \omega \delta\omega^2}{2\pi
q^2E_{rms}^2k_\parallel^2}\nonumber\\
\!\!&\approx&\!\!1.6\frac{10^3}{\omega}
\left(\frac{\delta\omega}{\omega}\right)^2
\left(\frac{\mathrm{A}}{\mathrm{Q}}\right)^2
\left(\frac{100\mathrm{V\;m}^{-1}}{E_{rms}}\right)^2
\times\nonumber\\
\!\!&\times&\!\!\left(\frac{1\mathrm{m}^{-1}}{k_\parallel}\right)
\left(\frac{\mathrm{B}_0}{0.01\;\mathrm{T}}\right)^4~.
\end{eqnarray}
With Equation ~\ref{heat_perp} the corresponding energy gain is
$v_\perp^2\approx\omega\delta\omega/k_\parallel^2$, which is only a
few $\mathrm{keV\;amu}^{-1}$ for the given parameters. It is too
little to explain the
observation, assuming realistic conditions.

%
%
\subsubsection{Bidirectional Propagation}
\label{bi}
Further acceleration is possible in the situation of the EFI where
counterpropagating electromagnetic waves of
the same sense of polarization are present. The basic idea is, that
the time average of the wave forces in parallel direction of
counterpropagating waves averages out to some extend. That way,
the ion can stay in resonance for much longer time.

In the following lefthand circular polarization is assumed.
Plus and minus signs in index positions therefore indicate the
direction of propagation and no longer sense of polarization.
The parallel acceleration now reads
\begin{eqnarray}
\dot{v}_\parallel &=& \sum_{j=1}^{N} \frac{q k_{\parallel,j}
v_\perp}{\omega_j m}\times\nonumber\\
&\times&\left[\stackrel{+}{e_j}\cos{\Phi_{+,j}}-\stackrel{-}{e_j}
\cos{\Phi_{-,j}}\right]~,
\label{acc_par_2}
\end{eqnarray}
where the phases $\Phi_{\pm,j}=\theta \mp k_{\parallel,j} v_\parallel t + \omega_j t$ are
introduced. Than $\Phi_{-,j}$ can be expressed by $\Phi_{+,j}$ and we rewrite
Equation~\ref{acc_par_2} as
\begin{eqnarray}
\dot{v}_\parallel &=& \sum_{j=1}^{N} \frac{q k_{\parallel,j}
v_\perp}{\omega_j m}
\left[(\stackrel{+}{e_j}-\stackrel{-}{e_j})\cos{\Phi_{+,j}}\right.\nonumber\\
&+&2\left.\stackrel{-}{e_j}\sin{\left(\theta +\omega_jt\right)}\sin{\left(k_{\parallel,j}
    v_\parallel t\right)}\right]~.
\label{acc_par_3}
\end{eqnarray}
The first term in the bracket is similar to the unidirectional case
(Eq.~\ref{acc_par}) except that the difference between the
bidirectional fields enters now and reduces the parallel
acceleration. As in the unidirectional case this term is nearly
constant in resonance, i.e. $\Phi_{+,j}\approx0$.
The second term in the bracket contains a product of two sinus
terms. At times $t_v<2\pi/k_{\parallel,j}v_\parallel$ there is a mean
change in $v_\parallel$ when averaging over $t$ due to the first sinus
term. The second sinus term, $\sin{\left(k_{\parallel,j}v_\parallel
    t\right)}$ leads to an oscillatory motion in $v_\parallel$ for
$t_v>2\pi/k_{\parallel,j}v_\parallel$ and thus slows down the escape
from the resonance region and, eventually,
allows the regain of resonance. The maximum
energy, that can be reached by the ion therefore can be increased. The
dynamics of $v_\parallel$ (Eq.~\ref{acc_par_3}) is nonlinear and
treated in numerical simulations with the exact wave spectrum and
dispersion relation of the counterpropagating waves.
%
%
\subsection{Test-Particle Simulation}
\label{sim}
\begin{figure*}[htb]
   \epsscale{2}
   \plottwo{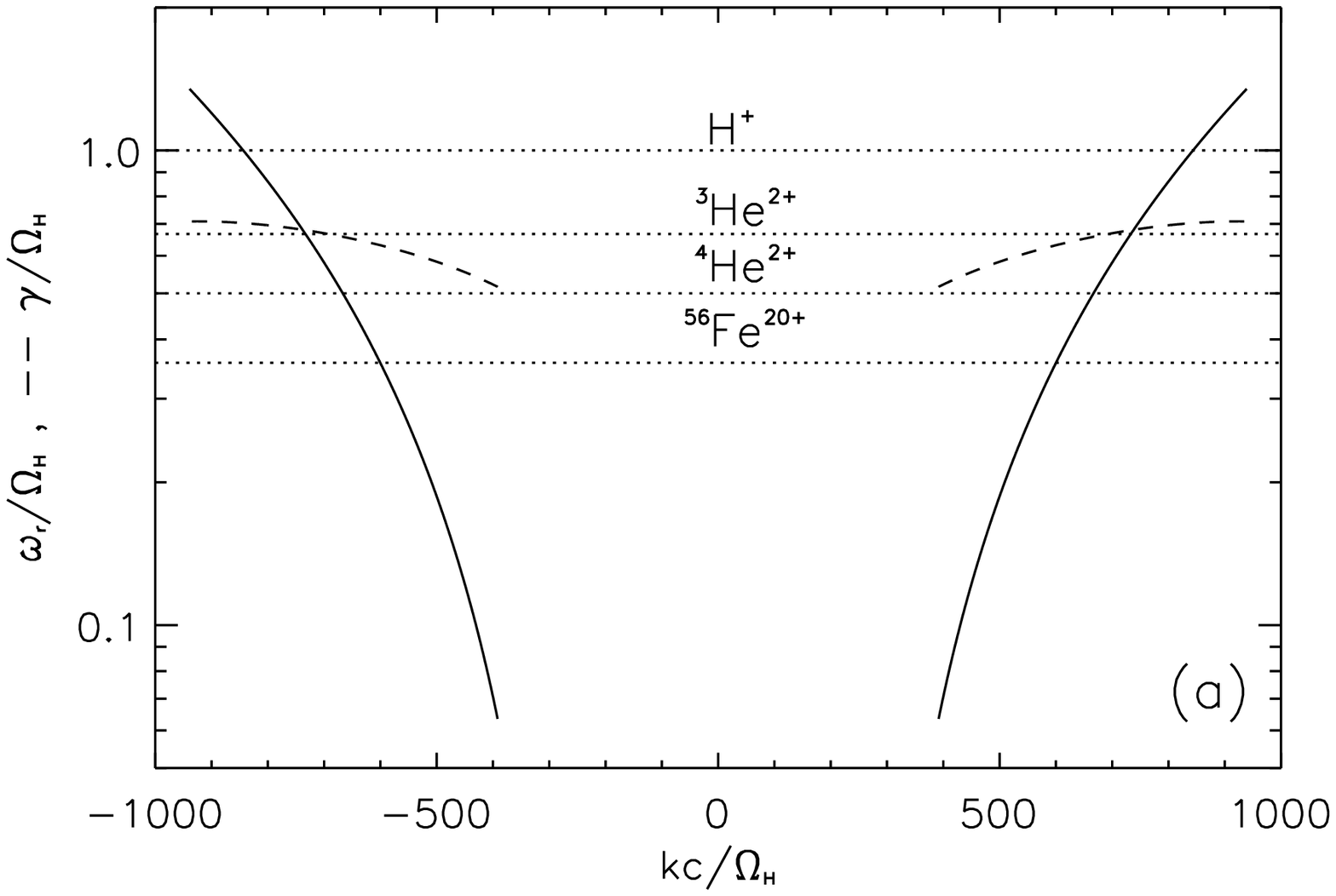}{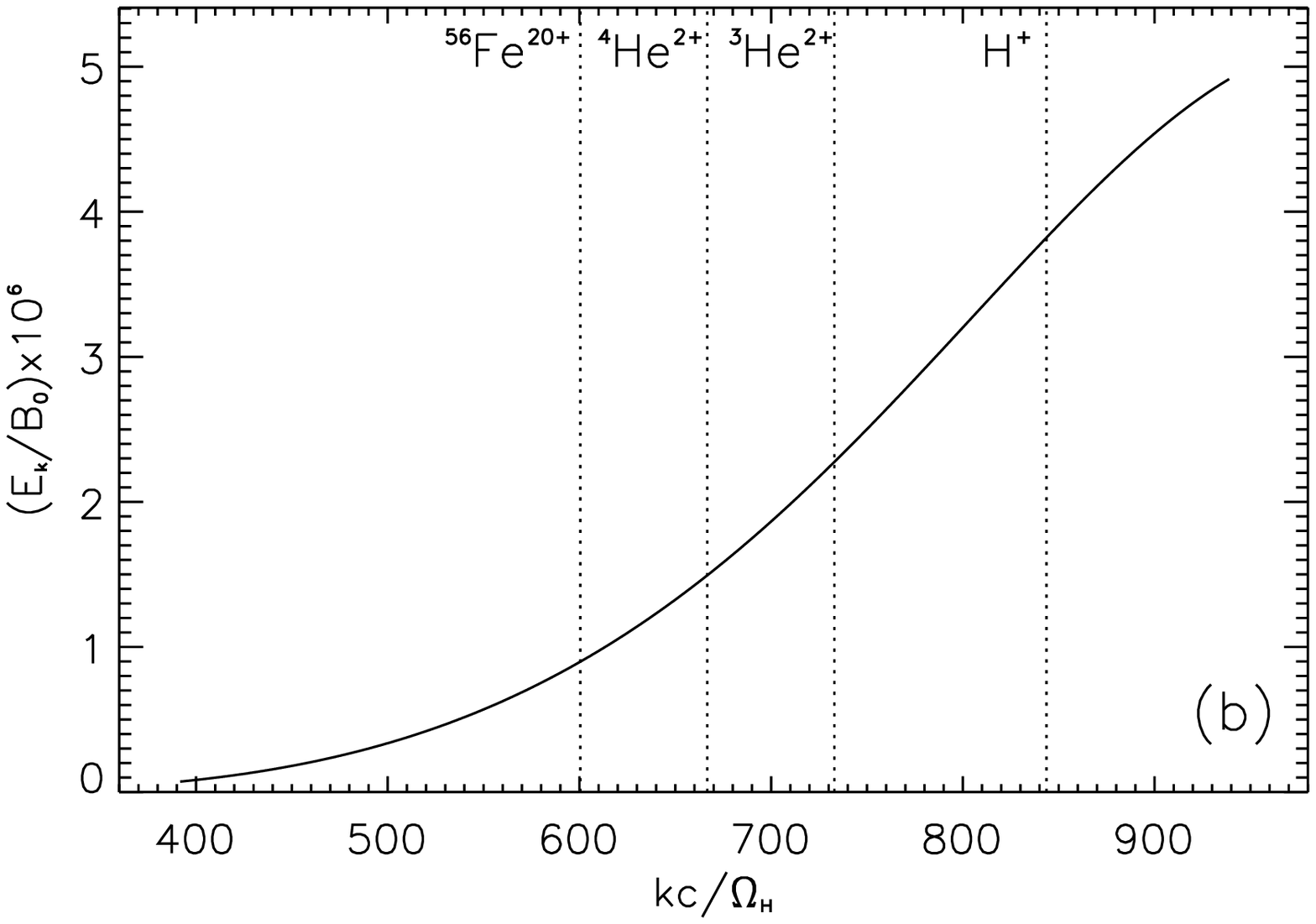}\\
   \plotone{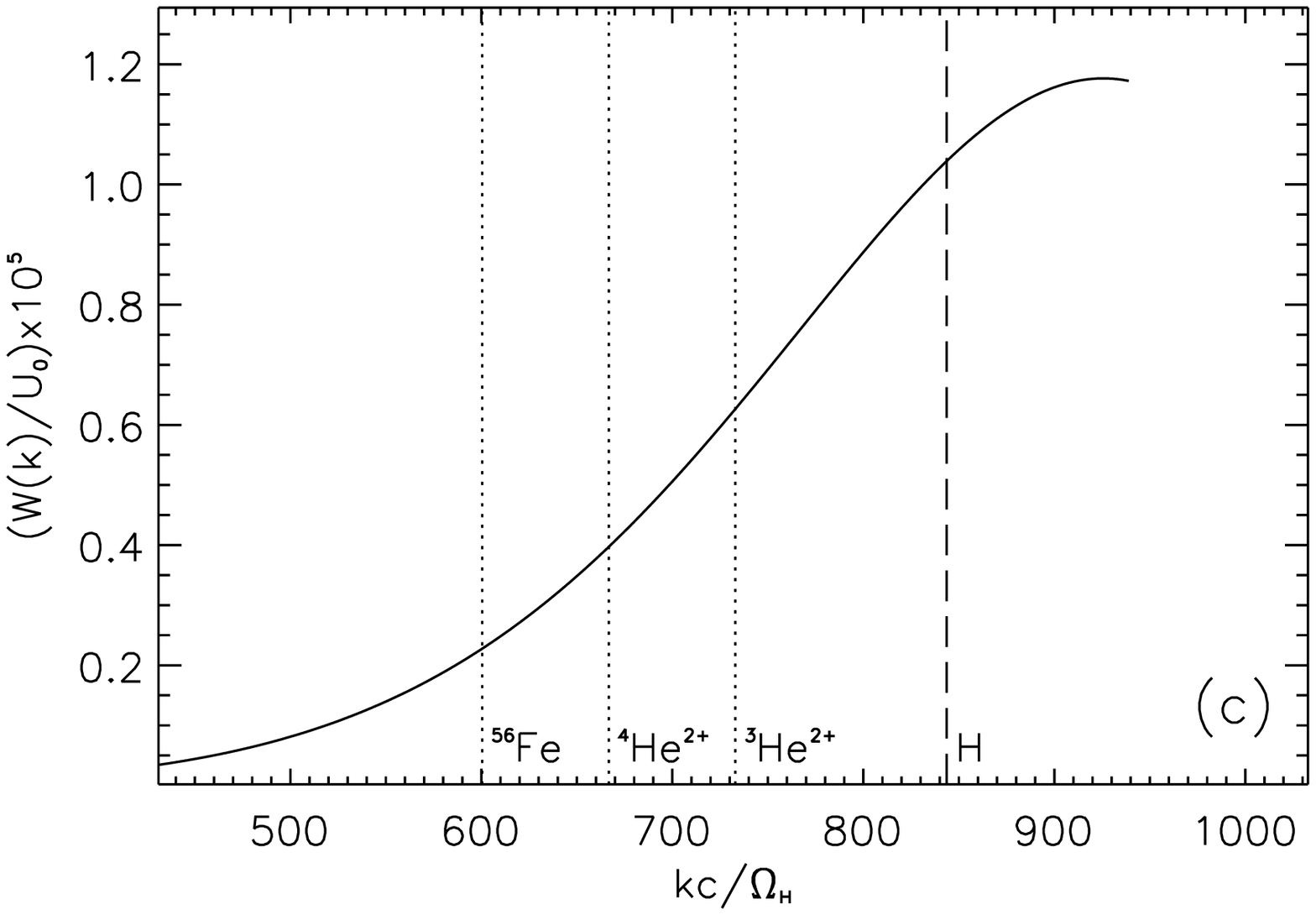}
 \caption[]{Properties of the wave spectrum assumed for the test
   particle simulations. {\em (a)} Dispersion of the wavefield used for the
   test-particle simulation obtained from linearized theory. Each
   branch (positive and negative $k$)
   consists of 1000 monochromatic waves.  The dotted lines indicate
   the resonance frequencies of the according ions.
   {\em (b)} The electric field amplitudes of the wave spectrum at the end of
   the growth time. The vertical dotted (dashed) lines refer to the k
   values corresponding to the according ion gyrofrequencies. {\em (c)} The
   corresponding spectral energy density.}
 \label{fig:4}
\end{figure*}
The relativistic equations of motion for a particle of charge $q$
and mass $m$ in a field of $N$ waves and a homogeneous background
magnetic field $B_0$ are given by
\begin{eqnarray}
\frac{d\mbox{\boldmath $p$}}{dt}&\!\!\!\!\!=&\!\!\!\!\!q
\frac{\mbox{\boldmath $v$}}{c}\times \mbox{\boldmath $B$}_0+
q\sum_{k=1}^N\left(\mbox{\boldmath $E$}_k+\frac{\mbox{\boldmath $v$}}{c}\times
\mbox{\boldmath $B$}_k\right)\label{momentum}\\
\frac{d\mbox{\boldmath $x$}}{dt}&\!\!\!\!\!=&\!\!\!\!\!
\mbox{\boldmath $v$}\quad,\label{space}
\end{eqnarray}
\begin{figure*}[htb]
 \plottwo{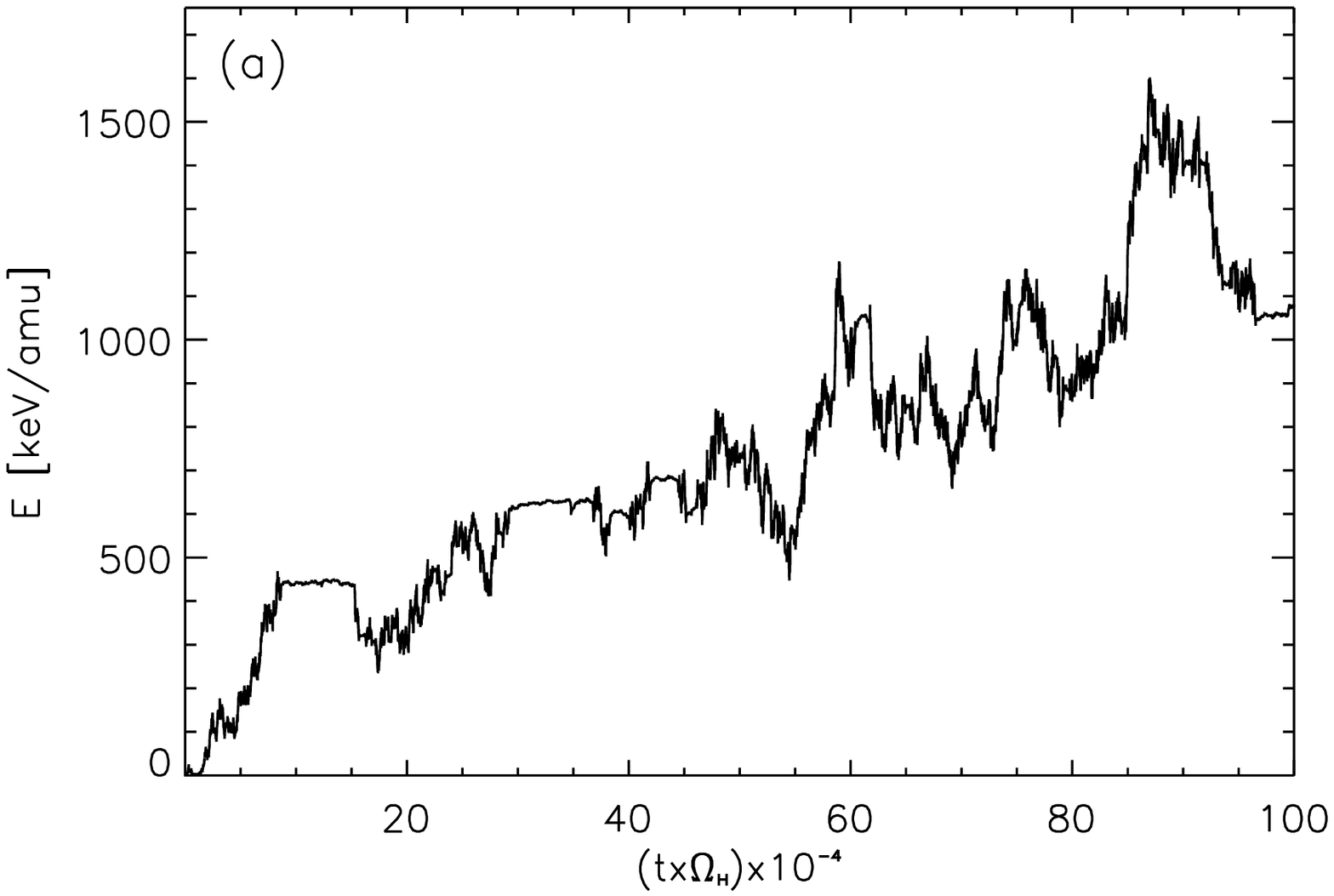}{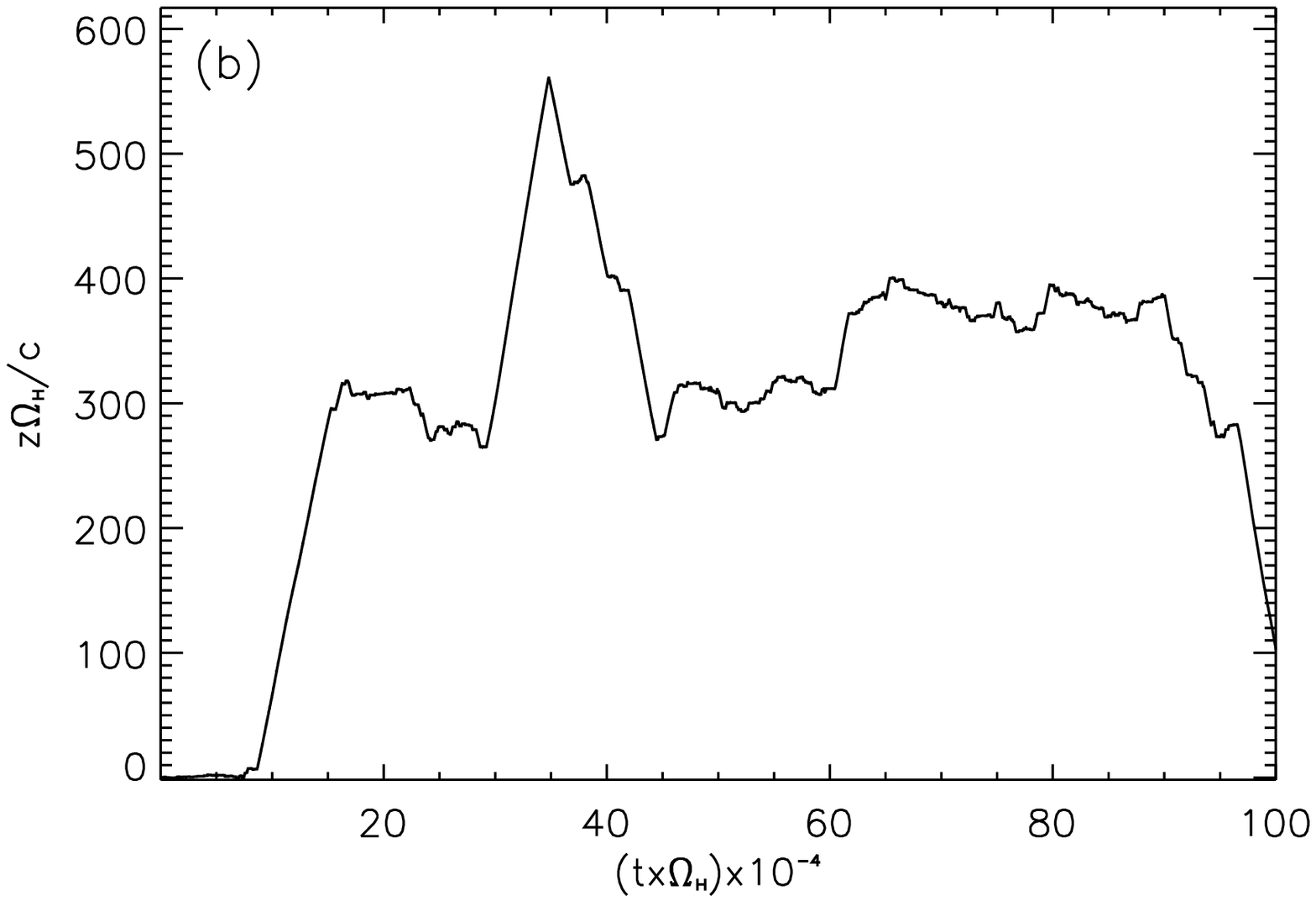}
   \epsscale{4.5}
 \plottwo{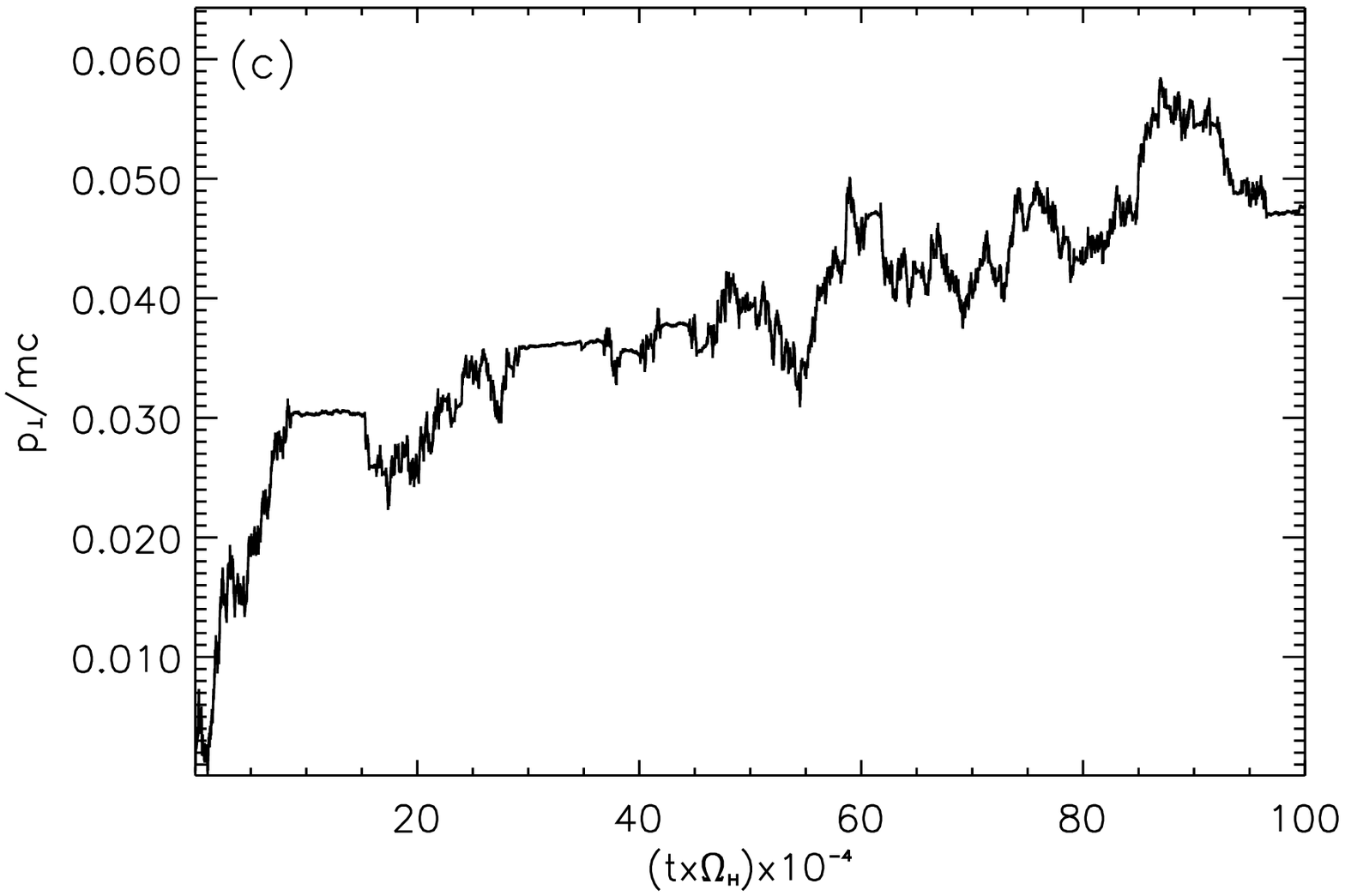}{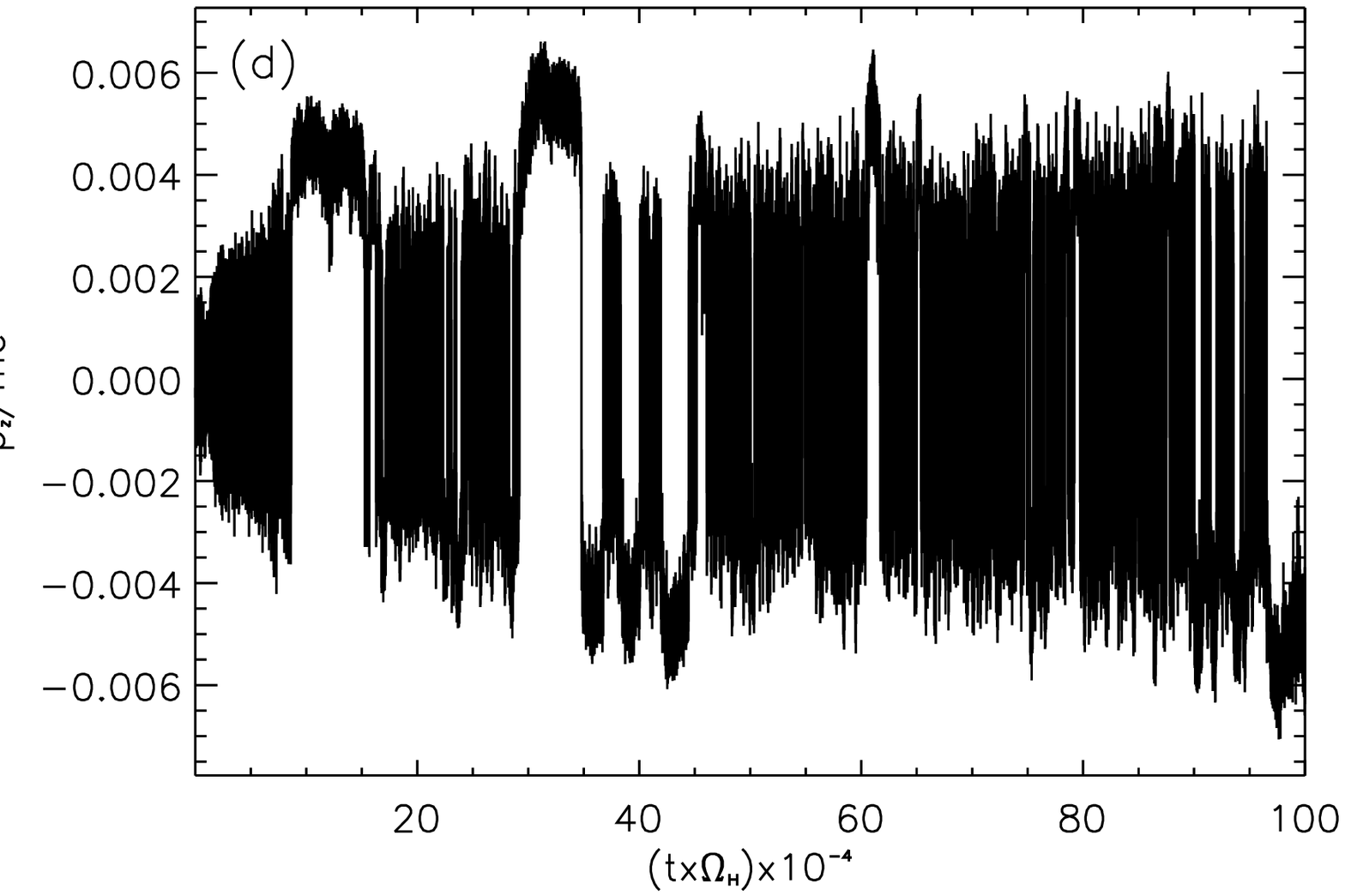}
 \epsscale{1}
 \caption[]{Temporal history of a typical \hedreiplus
 test-particle. {\em (a)} The kinetic energy in $\mathrm{keV\;amu}^{-1}$ of
 an initially thermal  \hedrei ion (100 $\mathrm{eV\;amu}^{-1}$) vs. time
 in units of $10^{4}\Omega_\mathrm{H}^{-1}$. {\em (b)} Distance traveled by
 the particle along the background magnetic field $B_0$. {\em (c)}
 Perpendicular and {\em (d)} parallel momentum of the particle vs. time.}
 \label{fig:5}
\end{figure*}
where $\mbox{\boldmath $x$}$ is the particle position vector,
$\mbox{\boldmath $v$}$ the velocity and $\mbox{\boldmath $E$}_k$ and
$\mbox{\boldmath $B$}_k$ are the electric and magnetic fields of the
wave $k$.

Test-Particle trajectories are calculated by integrating a
dimensionless form of Equations~(\ref{momentum}) and (\ref{space}) with a
standard leap-frog mover following \citet{birdsalllangdon1991}
(half acceleration with an
extrapolated electric field, rotation around the instantaneous
magnetic field, which includes the wavefield, and half
acceleration with updated velocities). The code used
herein is a leap-frog version of the code used by
\citet{millervinas1993} and has been tested against their version.
The simulation results of both codes are in excellent agreement on
timescales relevant for the analysis herein.

The time step in the simulation was
$\tau=0.1\Omega_\mathrm{H}^{-1}$.
The background magnetic field is considered to be homogeneous.
The wave frequencies, parallel wavenumbers and relative strength
of electric field components are obtained from the IDLWhamp
code described in \S~\ref{props}. The electric field strength
$E_k$ of one single wave $k$ is obtained  from growth according to
linear theory and the magnetic field components are calculated
according to Faraday's law. A wavefield consisting of 2000
monochromatic waves has been applied, confined to a range of
frequencies where the growth rate does not drop below $\sim 60\;\%$
of the maximum growth rate. All properties of the wavefield besides
the relative phases are symmetric with respect to the zero of
$k_\parallel$. The phases were chosen randomly for all waves
independent of the direction of propagation. Only waves propagating
parallel to the background magnetic field are considered.
Some properties of the applied wave spectrum are displayed in
Figure~\ref{fig:4}. The waves are purely transverse and lefthand
circularly polarized.

The spectral energy density of the wavefield is obtained by letting
the waves exponentially grow according to
$W(\tau,k)=W_0(k)\exp{\left(2\gamma(k)\cdot\tau\right)}$
from a thermal level $W_0(k)=k_BT$ up to an energy density of $W(\tau,k)$
at a time $\tau$ and $\gamma(k)$ is assumed to be
constant during this time. The time of growth has been chosen such, that
the total energy density in the wave field is
$W_\mathrm{tot}(\tau)/U_0=0.01$ at a time $\tau$, where $U_0=B_0^2/8\pi$ is the
energy density of the background magnetic field.

The plasma parameters are
$n_e=n_p=5\cdot10^{10}\;\mathrm{cm}^{-3},
T_\perp^e=T_\perp^p=1\cdot10^7\;K$,
$T_\parallel^e/T_\perp^e=15$, $T_\parallel^p/T_\perp^p=2$ (see Fig.~\ref{fig:1}) and the
background magnetic field is $B_0=100\;\mathrm{G}$.

To reduce computing time, the simulations have been divided in two
parts: {\em I)} For each of the two ion species, a small number of
particles have been followed for a long time
($\sim 1\;s$) to establish the ability of the  process to accelerate
ions to the requested energies (\mev). About 500
particles were computed in order to determine also the spread in
energy of the particle population. {\em II)} A larger number of ions (5000
particles for each species) have been followed for shorter times
$(\sim 0.052\;\mathrm{s})$ to
establish the energy distribution which then was applied to the
results of the first runs.

A representative run of the part {\em I)} is depicted in
Figure~\ref{fig:5}. The initial energy of the ion is
$100\;\mathrm{eV\;amu}^{-1}$, corresponding to a thermal velocity at a
temperature of $1\cdot10^7\;\mathrm{K}$, and the pitch angle cosine is
$\mu=0.5$. The heating rates in the simulation are not very
sensitive to variations in the initial energy of the ion. All
particles start at space coordinates equal to zero. For each particle
the phases of the wavefield have been chosen randomly. The total
energy in the wavefield is $U_\mathrm{W}/U_0\approx 0.01$
corresponding to the wave electric field amplitudes
$E_k/B_0$ depicted in Figure~\ref{fig:4}. The plasma parameters are chosen
as described in \S~\ref{idea}.

The typical energy evolution of a \hedrei ion shows intervals of zero
gain (Fig.~\ref{fig:5}{\em a)}. These are times when the ion looses
resonance with the wavefield due to a large excursion of
$v_\parallel$ from its oscillatory behavior described by
Equation~\ref{acc_par_3}. During these
intervals, the ion oscillates in the wavefield without being
significantly accelerated. Due to non-resonant interaction
with the waves, the ion is scattered back into resonance and can be
further accelerated. The loss and gain of resonance is
mirrored in panel {\em (b)} and {\em (d)} of Figure~\ref{fig:5}. During the
intervals of no resonance
(e.g. $(\mathrm{t}\times\Omega_\mathrm{H})\times10^{-4}\approx 8-15$), the ion
propagates freely and the
coordinate along $B_0$ (Fig.~\ref{fig:5}{\em b)} is just a straight line,
increasing or decreasing proportional with time. As can be seen in
Figure~\ref{fig:5}{\em d} the parallel momentum no longer oscillates around
zero in the above time interval. When the ion re-enters
resonance, the parallel momentum starts oscillating around zero,
indicating ping-pong trapping of the particle, and the coordinate along $B_0$
does not linearly change anymore.

\begin{figure}[htb]
 \epsscale{0.9}
 \plotone{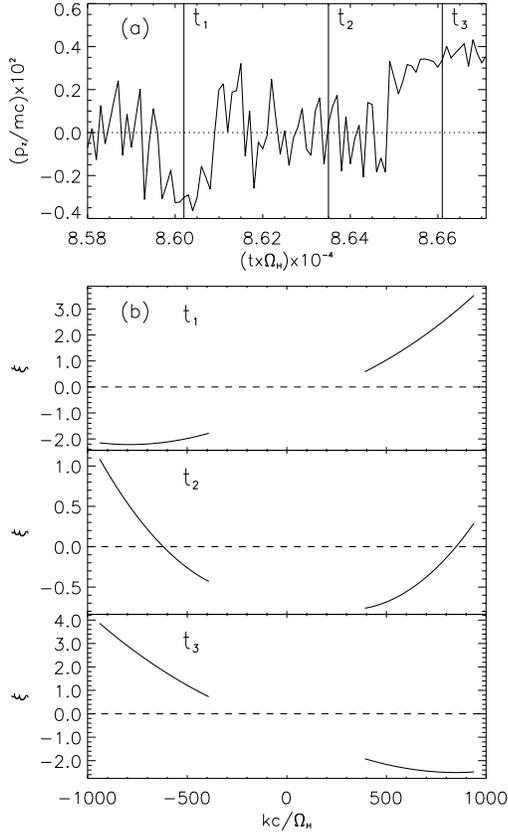}
 \epsscale{1}
 \caption[]{Zoomed region of the timeserie displayed in
 Figure~\ref{fig:5}{\em a}. Panel {\em (a)} shows the time interval
 8.58-8.67 in normalized time of the parallel momentum. {\em (b)}
 Plots of the frequency mismatch parameter
 $\xi=\omega_r-k_\parallel v_\parallel-\Omega_{3}$ at
 three different times as indicated in panel {\em (a)}.}
 \label{fig:6}
\end{figure}

To further illustrate the ping-pong behavior, the so-called
frequency mismatch parameter $\xi=\omega_r-k_\parallel
v_\parallel-\Omega_{3}$ (where $\Omega_3$ is the \hedreiplus
gyrofrequency) has been plotted in
Figure~\ref{fig:6}. Resonant interaction of the particle with a wave
means $\xi\approx 0$, reproducing the resonance
condition. Panel {\em (a)} of Figure~\ref{fig:6} displays a small
portion of the timeserie of Figure~\ref{fig:5}{\em a} and the three
vertical lines indicate times at which $\xi$ was computed. As can be
seen in Figure~\ref{fig:6}{\em b}, the particle is out of resonance at
times $t_1$ and $t_3$ while it is resonantly interacting with both
wave branches at time $t_2$. Accordingly, the parallel normalized
momentum $p_z/mc$ does not exhibit any oscillatory behavior
around zero at times $t_1$ and $t_3$ in
Figure~\ref{fig:6}{\em a} but clearly oscillates around
zero at $t_2$.

Acceleration is clearly very efficient with an averaged systematic
energy gain $\langle dE/dt\rangle$ of $\approx1.2\;\mathrm{MeV}\;
\mathrm{amu}^{-1}\;\mathrm{s}^{-1}$ (Fig.~\ref{fig:5}{\em a}). A
linear energy increase is predicted by Equation~\ref{heat_perp}. The
paths of more than 500 \hedrei ions have been simulated and exhibit
similar energization.
%
%
\section{ABUNDANCE ENHANCEMENTS}
\label{enhancement}
Among the overabundant ions observed during impulsive solar flares,
\hedrei takes a special position. Whereas most ions exhibit an
enhancement by factors of 1-10 with respect to coronal values, \hedrei
shows an excess of $\sim 2000$. Since it has been established that the
selective enhancement of ion abundances correlates with the
charge-to-mass ratio \citep{reames1998} and hence the gyrofrequency
of the according ions,
resonant wave-particle interaction is generally
believed to be the accelerating mechanism.
This led earlier theories to the approach of finding
special plasma waves existing only in a narrow range in the vicinity
of the \hedrei gyrofrequency. In order to generate the appropriate waves
(e.g. EIC waves \citep{fisk1978},
$\mathrm{H}^+$~EMIC waves in \citet{temerinroth1992}) peculiar plasma
conditions
have to be postulated. In the first case a pre-flare enhancement of
\hevier over $\mathrm{H}$ and in the latter case dense low-energy beams have to
be assumed in order to excite the waves in the appropriate frequency
ranges.

A different approach is investigated in the work presented
herein. EF waves, excited by an anisotropic velocity distribution
function resulting from bulk acceleration of electrons, accelerate
\hedrei and \hevier. Selectivity is
achieved by an energy density profile of the waves self-consistently
produced from linear growth.
For nonresonant instabilities, this profile is approximately true not
only during the growth 
phase but also for the saturated state of the instability (see \S~\ref{idea}).
As can be seen from Figure~\ref{fig:4} the
resulting energy density profile clearly supports an enhanced
acceleration of \hedrei above $^4\mathrm{He}$. The energy density in the wave
field is higher at the gyrofrequency of \hedrei than
$^4\mathrm{He}$. More
energy is therefore available for the acceleration of $^3\mathrm{He}$.
The quantitative heating rates of an ensemble of
ions of the two species have been determined by numerical
simulations where the total energy density in the wavefield has been
treated as a free parameter.

In order to compare the simulation results to the observational values
of \heratio, abundances have to be compared at the same
energies. Observations from the {\em ISEE 3} spacecraft were
taken in the $1.3-1.6\;\mathrm{MeV\;amu}^{-1}$
channel~\citep{reamesetal1994} and observations  by
\citet{moebiusetal1982} with the {\em ISEE 1} and {\em ISEE 3}
spacecrafts were taken in the $0.4-4.0\;\mathrm{MeV\;amu}^{-1}$
range. According to these measurements, a viable mechanism for \hedrei
acceleration and its enhancement above \hevier has to meet the
requirements of \S~\ref{idea}: {\em (i)} Ion energies $\sim
1\;\mathrm{MeV\;amu}^{-1}$ on timescales of $1\;s$, {\em (ii)}
\heratio$>0.1$ above \mev, {\em (iii)} a total of about $10^{31}$
$^3\mathrm{He}$ nuclei above $1\;\mathrm{MeV\;amu}^{-1}$, {\em (iv)} \hedrei
spectrum harder than the \hevier spectrum in the range of
$0.4-4.0\;\mathrm{MeV\;amu}^{-1}$ and {\em (v)} a turnover in the particle
energy spectrum of \hedrei at around $\sim100\;\mathrm{keV\;amu}^{-1}$.

In the following, the presented model is investigated in view of these
requirements.

\begin{figure}[ht]
 \plotone{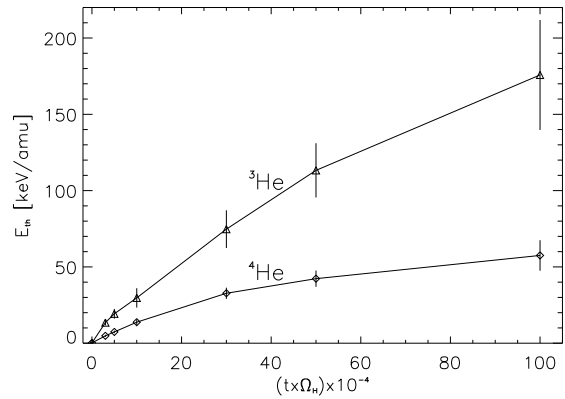}
 \caption[]{Temporal evolution of thermal energy $E_{th}(t)$ as defined in
   Equation~\ref{maxwell_energy}. A maxwellian distribution has been
   fitted to an ensemble of 500 particles for each species.}
 \label{fig:7}
\end{figure}
The ability of the EF waves to accelerate the ions to the relevant
energies has been shown in \S~\ref{acc} and requirement {\em (i)} therefore
is fulfilled. In order to address items {\em (ii)} - {\em (v)} the collective
behavior of an ensemble of ions has to be investigated.

From part {\em II)}  of the numerical simulations (5000 particles,
short times) it is found 
that the population of ions in energy develops according to a
maxwellian distribution function 
for a system of three degrees of freedom
\begin{eqnarray}
\label{maxwell_energy}
f_\alpha(E,t)= \frac{n_\alpha}{\sqrt{2\pi
    E_\alpha^{th}(t)^3}}\sqrt{E}\exp{\left(-\frac{E}
    {2E_\alpha^{th}(t)}\right)},
\end{eqnarray}
where $f_\alpha(E,t)$ has been normalized to $n_\alpha$, the density
of species $\alpha$.

By fitting this distribution model to the runs of
part {\em I)} (500 particles, long times) the thermal energy in time
$E_\alpha^{th}(t)=1/2\;\mathrm{m}_\alpha (v_\alpha^{th}(t))^2$, and
hence the heating rates, were extracted. The function
$E_\alpha^{th}(t)$ for \hedrei and \hevier is depicted in
Figure~\ref{fig:7}. Clearly, \hedrei is accelerated faster than
\hevier and this causes an enhancement at a given energy (say
$1\;\mathrm{MeV\;amu}^{-1}$).

\begin{figure}[ht]
 \plotone{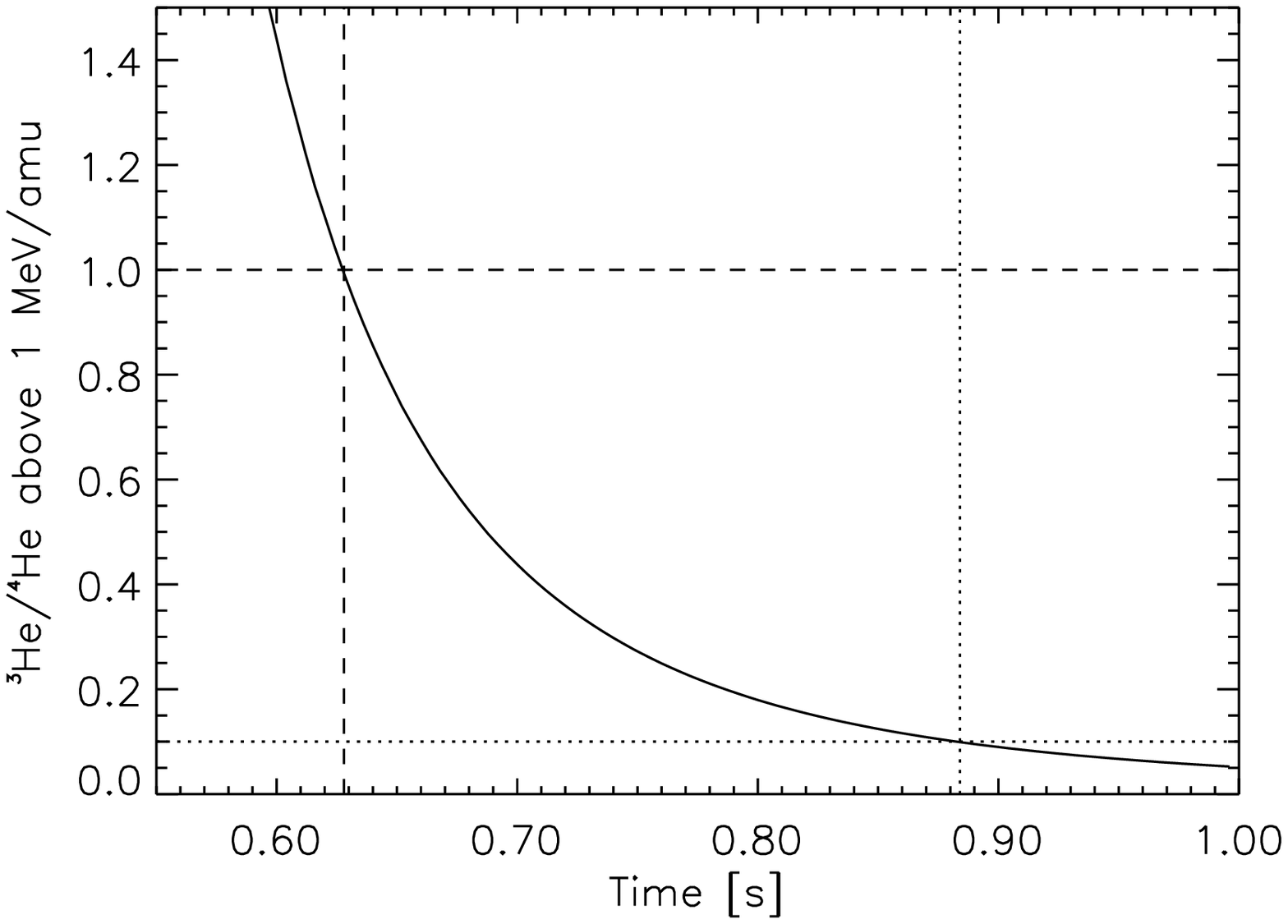}
 \caption[]{Abundance ratio \heratio of ions above \mev vs. acceleration time.}
 \label{fig:8}
\end{figure}
The ratio \heratio above \mev can now be calculated
in dependence on acceleration time. We integrate
\begin{eqnarray}\label{density_above}
n_\alpha(t)=\int\limits^\infty\limits_{1\;\mathrm{MeV\;amu}^{-1}}
\!\!\!\!\!\!\!\!\!\!f_\alpha(E,t)\;dE\quad,
\end{eqnarray}
where $\alpha=3,4$ corresponds to the according He isotope,
and form the ratio
$(^3\mathrm{He}/^4\mathrm{He})(t)=n_3(t)/n_4(t)$. It is found that the
required abundance ratio of $^3\mathrm{He}/^4\mathrm{He}\sim0.1-1$ is
reached in the time
range of $t_0\sim0.63-0.89\;s$ (see Fig.~\ref{fig:8}). Due to the
higher acceleration rate of \hedrei (about factor of 2--3 above
$^4\mathrm{He}$),
the fast tail of the energy
distribution function populates energies above \mev
faster than for $^4\mathrm{He}$. This yields a \heratio$\gg1$ for small times
$t\ll t_0$, \heratio$\sim0.1-1$ for $t_0\sim 0.63-0.89\;\mathrm{s}$
and asymptotically approaches the
coronal abundance ratio for times $t\gg t_0$.
The derived time range of $t_0$ corresponds well to the typical
timescale of electron acceleration during the impulsive phases of
flares of $0.1-1\;s$ as manifest by the shortest hard X-ray
peaks~\citep{kiplingeretal1984}. Since the EF wavefield is a direct consequence
of the electron acceleration process, the lifetime of the field is
expected to be on the same order of magnitude. When electron
acceleration ceases, the \heratio ratio is frozen and no further
ion acceleration occurs.

The time integrated total number of \hedrei nuclei being accelerated
above $1.3\;\mathrm{MeV\;amu}^{-1}$ in an impulsive flare is about
$10^{31}$ particles~\citep{reamesetal1994}. Assuming the flaring area
to be of about $10^{17}\;\mathrm{cm}^2$ with a scale height of
$10^9\;\mathrm{cm}$, the proton density to be
$n_H=5\times10^{10}\;\mathrm{cm}^{-3}$ and
$^3\mathrm{He}/\mathrm{H}=5\times10^{-5}$, there are $2.5\times
10^{32}$ \hedrei ions available in the flaring volume. This yields a
percentage of about $4\%$ of the pre-flaring \hedrei population
that has to be accelerated above $1\;\mathrm{MeV\;amu}^{-1}$. Computing
the value of the \hedrei density $n_3(t_0)$ at time $t_0$ and
comparing it to the coronal
abundance yields a percentage of about $\sim3.5-10.5\%$. Thus the number of
accelerated ions at the required energies is sufficient to explain the
observed particle fluxes. The model therefore also accounts for
criterion {\em (iii)}.

\begin{figure}[ht]
 \plotone{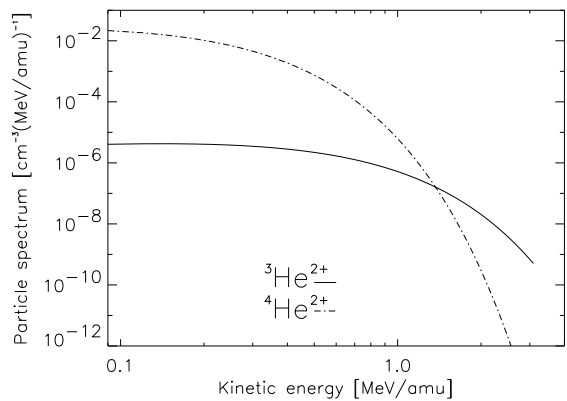}
 \caption[]{Particle energy spectrum in $\mathrm{MeV\;amu}^{-1}$ at time
   $t_0=0.76\;\mathrm{s}$. \hedrei is harder than
   \hevier, qualitatively reproducing the spectrum observed by
   \citet{moebiusetal1982}.}
 \label{fig:9}
\end{figure}
The \hedrei acceleration by EF waves  also generates ion energy
spectra that qualitatively reproduce the observed behavior. According
to \citet{moebiusetal1982} the observed spectrum of \hedrei
is harder than the spectrum of \hevier in the energy range of
$0.4-4.0\;\mathrm{MeV\;amu}^{-1}$ which
is a feature of all $^3\mathrm{He}$ rich periods in their study.
The according particle densities in dependence on
energy per nucleon resulting from the model presented herein are
depicted in Figure~\ref{fig:9}, reproducing the observations.

A turnover of the \hedrei spectrum at around $\sim 120\;\mathrm{keV\;amu}^{-1}$
energies can be seen in Figure~\ref{fig:9}. This reproduces very
nicely the observations by
\citet{masonetal2000} who reported a turnover at around
$\sim100\;\mathrm{keV\;amu}^{-1}$. It can be interpreted as the result of
the whole \hedrei population being accelerated by EF waves.

Therefore it has been established that the EFI is not only a viable
acceleration mechanism for ions during solar flares, it also can
account for the observed abundance enhancements of \hedrei over \hevier,
accelerates enough $^3\mathrm{He}$ nuclei, and reproduces the
qualitative behavior of the particle energy spectra.
%
%
\section{PROTONS AND HEAVIER IONS}
\label{heavy}
The energy density profile may suggest by the same arguments used
to explain the enhancement of \hedrei above $^4\mathrm{He}$, that
$\mathrm{H}$ should be
enhanced above $^3\mathrm{He}$, which is not the case. This is not a
real problem of the model because of the
very different number densities of the according ion species. The energy
in the wavefield available for proton acceleration is higher than the
energy available for $^3\mathrm{He}$. Nevertheless, there are much more protons.
Thus the available energy is distributed among a larger number of
nuclei and the energy gained per nucleon is smaller for the protons.
By the same argument, \hedrei should be further enhanced in comparison
to $^4\mathrm{He}$. This effect was neglected herein
since the scope of this paper is to show the ability of fractionation
by the accelerator itself.

When discussing the possible acceleration of the protons by EF waves,
other effects have to be taken into account. Whereas, due to their low
densities, \hevier and \hedrei can
be treated as test-particles, this view is not valid for protons and
nonlinear effects as the erosion of the spectral wave energy around
the gyrofrequency of $\mathrm{H}$ have to be considered. Such
an analysis requires simulations that are beyond the scope of the
presented work.

\hedrei rich events usually are also characterized by enhanced
\feratio ratios. Although the observed enhancement of \feratio is only
about a factor
of 8 over the coronal ratio, it is still the strongest
enrichment of the heavier ions. Other heavier ions as Ne, Mg, Si are
enhanced by factors of 2--3 above O, whereas C, O, N are
not enhanced. In this section the enrichment of Fe ions is
addressed and briefly discussed in view of the model presented in this
work.

In a plasma at a temperature of $10^7\;\mathrm{K}$, the most probable charge
state of $^{56}\mathrm{Fe}$ is 20+ while $^{16}\mathrm{O}$ is fully
ionized. When resonantly cyclotron accelerated, \ox should exhibit the
same acceleration rate as \hevier while \fe, following the
argumentation for enhanced \hedrei acceleration rate, is much slower heated.
As can be seen from Figure~\ref{fig:4}{\em c} the energy available for
\fe is much smaller than for \ox. This behavior was confirmed by
carrying out the same simulations for \fe and \ox as for \hedrei and
\hevier. Although the acceleration via EF waves by cyclotron resonance
energizes \fe ions up to energies of \mev on timescales of
$\sim10\;\mathrm{s}$, an
enhancement of \feratio by the same mechanism as for \heratio
presented above cannot be reached. It is in general impossible in
the EFI model to explain a simultaneous enhancement of \heratio and
\feratio when the spectral energy density exhibits a slope as depicted
in Figure~\ref{fig:4}{\em c}. The same energy is available for
\hevier and \ox, but \hedrei has more and \fe has less energy available.
The heating rate of \fe will therefore always lie below the \ox and
\hevier rate, whereas \hedrei always lies above it in the present
simple model. Other processes may enter the picture. Non-linear
effects as the back reaction of the wavefield in
response to energy loss to the ions can have a significant influence
on the heating rates. However, the analysis of such effects are beyond
the scope of this paper and will not be addressed here.

Similar problems arise for the enrichment of Ne, Mg, Si since their
gyrofrequencies also lie below the gyrofrequency of \hevier.
The same arguments as for \fe therefore make an enhancement above \ox
impossible in the present scenario.

The observed heavy ion enrichment is independent of the degree of \hedrei
enhancement, although heavier ion enrichments are generally associated with
\hedrei \citep{masonetal2000}. This is an indication that it may not be
the same mechanism that accounts for the acceleration of \hedrei and
\fe or other heavier ions.
%
%
\section{CONCLUSION}
\label{conc}
A model for the enrichment of \hedrei during impulsive solar flares is
presented. The acceleration of \hedrei and \hevier can be understood as
consequence of Electron Firehose (EF) waves, resulting from an
unstable anisotropic electron distribution with $T^e_\parallel>T^e_\perp$.
The model does not need additional sources of free
energy or plasma properties than those resulting directly from the
acceleration of the bulk of electrons.
The essential result of the investigation is that EF waves, excited by an
anisotropic electron distribution function, can accelerate \hedrei and
\hevier via cyclotron resonance to $\mathrm{MeV\;amu}^{-1}$ energies
on timescales of $\sim 1\;\mathrm{s}$. In support of this conclusion, the
following specific results are found:

1. The EF waves accelerate \hedrei ions via cyclotron resonance. The
  symmetry of EF waves in $k_\parallel$ greatly enhances the efficiency of the
  acceleration process with respect to the standard cyclotron resonant
  wave-particle interaction of a unidirectional propagating
  wavefield. The parallel force driving the ion out of resonance, and
  therefore limiting the acceleration time in
  the unidirectional case, cancels out in the time average for the
  counterpropagating wavefield. The total resonant
  interaction time therefore is strongly increased and the particle
  can reach high energies.

2. The linear growth of the EF waves self-consistently
  generates a spectral energy distribution, causing enhanced
  acceleration of \hedrei with respect to \hevier. It is found from
  test-particle simulations that the heating rate of \hedrei exceeds the
  values for \hevier by about a factor of 2--3.

3. It is shown that this small enhancement in the heating rate of
  \hedrei above \hevier already can account for the
  observed enhancement in \heratio from the coronal value of
  $\sim5\times10^{-4}$ up to $0.1-1$ during impulsive solar
  flares. Due to the larger heating rates, the high energy tail of
  \hedrei populates energies above \mev faster. It
  reaches the required abundance ratio of \heratio for an
  acceleration timescale of $\sim0.76\;\mathrm{s}$.

4. The fraction of \hedrei accelerated by the EF waves is large enough
  to account for the observed particle fluxes. Between $4-11\%$
  of the coronal \hedrei population in the flare is accelerated above
  \mev.

5. The acceleration model reproduces the qualitative behavior of the
  \hedrei energy spectrum with respect to the \hevier energy
  spectrum. As observed by \citet{moebiusetal1982} the
  \hedrei spectrum is generally harder than the \hevier
  spectrum. Moreover, a turnover in the \hedrei energy spectrum around
  \kev is obtained, reproducing the observations by
  \citet{masonetal2000}.

\smallskip

EF waves are an inherent property of stochastic electron acceleration
by transit-time damping of cascading fast-mode waves. Thus, their
ability of accelerating \hedrei and \hevier supports the scenario of
stochastic acceleration of the bulk electrons in impulsive
flares. 
Future work will have to extend these results to the oblique
EF mode and investigate its role in the enhancement of heavy ions and
in the flare acceleration process in general.

\acknowledgements
The authors thank Kjell R\"onnmark for providing them with a copy of the
KGI report describing the original WHAMP code. The authors also want
to acknowledge G\'erard Belmont and Laurence Rezeau who gave them free
access to their improved version of the WHAMP code, which has become
the mathematical core of IDLWhamp \citep{paesold2002}.

This work was financially supported by the Swiss National Science
Foundation (grant No. 2000--061559).
%
%

\end{document}